\documentclass[11pt]{article}

\def\genf#1#2{\buildrel#2\over{\mathfrak#1}}

\usepackage{epsfig}

\usepackage{amssymb}

\usepackage{amsmath}

\usepackage{latexsym}

\def\W{\wedge}
\def\D{\mbox{d}}

\def\slash#1{\, /\kern-0.6em{#1}}

\def\genf#1#2{\buildrel#2\over{\mathfrak #1}}

\def\gend{\mathbf d}

\begin{document}
\begin{center}
\vskip 0.0cm
\large{\bf Generalized vector field\footnote{Talk delivered  by
   S.Chatterjee at at 17th DAE-BRNS High Energy Physics Symposium
   (HEP06), Kharagpur IIT,
India, 11-16 Dec 2006}}  
\vskip 0.0cm {\normalsize Saikat Chatterjee, Amitabha
 Lahiri}\\{\normalsize{S.N.Bose National Centre for
 Basic sciences\\
Block-JD, Sector-III, Salt Lake\\Kolkata-700098\\}}

\end{center}

\begin{abstract}
We define generalized vector fields, and contraction and Lie
derivatives with respect to them. Generalized commutators are also
defined.
\end{abstract}


\section{\label{intro}Introduction}
The idea of a form of negative degree was first introduced by
Sparling~\cite{Sparling:avrc97, Perjes:esi98}. Nurowski and
Robinson~\cite{nurob:cqgl01, nurob:cqg02} took this idea and used
it to develop a structure of generalized differential forms. A
generalized $p$-form is an ordered pair of an ordinary $p$-form and
a $p+1$-form, with the wedge product of $\genf a
p\,=(\alpha_p,\,\alpha_{p+1})$ and $\genf b q
\,=(\beta_q,\,\beta_{q+1})$ being defined as $\genf a p \W \genf b
q = (\alpha_p\beta_q,\, \alpha_p\beta_{q+1} + (-1)^q \alpha_{p+1}
\beta_q)$, where $\alpha_{p}$ is an ordinary $p$-form, etc. The
exterior derivative is defined as $\gend \genf{a}{p}\, =
(\D\alpha_p+(-1)^{p+1}k\alpha_{p+1},\,\D\alpha_{p+1})$.  This
structure was expanded to include generalized vector fields defined
as an ordered pair of of ordinary vector and scalar fields
\cite{Generalizedvector}.  Here we discuss various geometric
operations such as contraction, Lie derivative, commutator etc. of
generalized vector fields.

\section{\label{genv}Generalized vectors and contraction}
Following \cite{Generalizedvector} we define a generalized vector
field as an ordered pair of an ordinary vector field $v_1$ and an
ordinary scalar field $v_0$,
\begin{equation}
V := (v_1, v_0).
\label{genv.def}  
\end{equation}
Clearly, the submodule $v_{0}=0$ of generalized vector fields can
be identified with the module of ordinary vector fields on the
manifold. Generalized scalar multiplication by a generalized
zero-form $\genf a0\, = (\alpha_0, \alpha_1)$ is defined as
\begin{equation}
\genf a0 V = (\alpha_0 v_1, \alpha_0 v_0 + i_{v_1}\alpha_1).
\label{genv.gscm}
\end{equation}
This is a linear operation, and satisfies $\genf a0 (\genf b0 V) =
(\genf a0 \W \genf b0) V.$ The interior product $I_V$ is defined
as  
\begin{equation}
I_V \genf{a}{p}\, =
(i_{v_{1}}\alpha_p\,, i_{v_{1}}\alpha_{p+1} +
p(-1)^{p-1}v_{0}\alpha_p)\,.
\label{genv.contract}
\end{equation}
This  satisfies Leibniz rule,
\begin{equation}
I_V (\genf{a}{p} \W \genf{b}{q}) =
(I_V \genf{a}{p})\W \genf{b}{q} +
(-1)^{p}\genf{a}{p} \W (I_V \genf{b}{q}),
\label{genv.antider}
\end{equation}
but  is linear only under ordinary scalar multiplication,
\begin{equation}
I_{V+\mu W}= I_V +\mu I_W,  
\label{genv.linear}
\end{equation}
where $\mu$ is an ordinary scalar field.

\section{\label{lieform}Lie derivative}
Equipped with the generalized exterior derivative and interior
product we can define the Lie derivative using Cartan's formula. We
will find that the resulting derivative is problematic when applied
on a generalized vector field and we have to add an extra
correction term. For the moment, let us define the generalized Lie
derivative $\mathcal L_V$ with respect to $V$ as,
\begin{equation}
\mathcal L_V\genf a p = I_V\gend\genf a p +
\gend I_V\genf a p\,.
\label{lieform.def}
\end{equation}
Since we know how to calculate the right hand side, we find
\begin{eqnarray}
\mathcal L_V\genf a p = (L_{v_1}\alpha_p - pkv_0\alpha_p\,, &&
L_{v_1}\alpha_{p+1} - (p+1)kv_0\alpha_{p+1} \, \nonumber \\
&& + p(-1)^{p-1}(\D v_0)\alpha_p + (-1)^p v_0 \D\alpha_p )\,,
\label{lieform.action}
\end{eqnarray}
where as usual $\genf a p\, =(\alpha_p,\alpha_{p+1})\,,
V=(v_1,v_0)\,,$ and $L_{v_1}$ is the ordinary Lie derivative with
respect to the ordinary vector field $v_1$.



To find the definition for the Lie derivative of a vector field, we
demand that the following equality holds for any two generalized
vector fields $V, W\,,$ and any generalized $p$-form $\genf a p\,$:
\begin{eqnarray}
\mathcal L_V (I_W \genf a p) = I_W(\mathcal L_V\genf a p) +
I_{(\mathcal L_V\,W)}\genf a p \,,
\label{lievector.Leibnitz}
\end{eqnarray}
where we have written $\mathcal L_V W$ for the action of $\mathcal
L_V$ on $W$. This is what we would like to define as the Lie
derivative of $W$ with respect to $V$.  Using
Eqs.~(\ref{genv.contract}) and (\ref{lieform.action}) however we find that
\begin{equation}
\mathcal L_V (I_W \genf a p) - I_W \mathcal L_V \genf a p =
I_{([v_1,w_1] + k v_0 w_1, L_{v_1}w_0 - L_{w_1}v_0)} \genf a p -
(-1)^p(0, L_{v_0w_1}\alpha_{p}),
\label{lievector.failure}
\end{equation}
which is not a contraction. This problem can be resolved
\cite{Generalizedvector} by modifying the formula for the Lie
derivative of a generalized $p$-form to
\begin{eqnarray}
{\widehat {\cal L}_V}\genf ap &=& \mathcal L_V\genf ap +
(-1)^p(0, -v_0 \D\alpha_p + p \D v_0 \alpha_p) \, \nonumber \\
&=& (L_{v_1}\alpha_{p}-pk v_0\alpha_{p}, L_{v_1}\alpha_{p+1} -
(p+1)kv_0 \alpha_{p+1})\,.
\label{lievector.lieform}
\end{eqnarray}
This new and improved generalized Lie derivative satisfies the
Leibniz rule,
\begin{equation}
{\widehat {\cal L}_V}(\genf ap \W \genf bq) = ({\widehat {\cal
L}_V}\genf ap) \W \genf bq + \genf ap \W ({\widehat {\cal L}_V}
\genf bq).
\label{lievector.leibniz}
\end{equation}
With this modified Lie derivative we find
\begin{equation}
{\widehat {\cal L}_{V}} I_W - I_W {\widehat {\cal L} _{V}}
= I_{([v_1,w_1] + kv_0w_1, L_{v_1}w_0)}\,.
\label{lievector.derivation}
\end{equation}
Therefore  the generalized Lie derivative of a
generalized vector field  is
\begin{equation}
{\widehat {\cal L}_V}W = ([v_1,w_1] + k v_0 w_1, L_{v_1}w_0).
\label{lievector.lievector}
\end{equation}
The commutator of two generalized Lie derivatives is also a generalized
Lie derivative itself,
\begin{equation}
{\widehat {\cal L}_V}{\widehat {\cal L}_W} - {\widehat {\cal L}_W}{\widehat {\cal
L}_V}   = {\widehat {\cal L}_{\{V,W\}}}\,,
\label{lievector.comm}
\end{equation}
which allows us define the generalized commutator as
\begin{eqnarray}
\{V,W\} = \Big([v_1\,, w_1]\,, L_{v_1}w_0 - L_{w_1}v_0 \Big)
\,.
\label{lieform.gencomm}
\end{eqnarray}
This commutator $\{V,W\}$ is antisymmetric in $V$ and $W$, bilinear  and  satisfies the Jacobi identity. For $U,V,W\in {\cal X}_G(M)\,,$ we find that
\begin{equation}
\{U,\{V,W\}\}+\{V,\{W,U\}\}+\{W,\{U,V\}\}=0\,.
\label{lieform.Jacobi}
\end{equation}
Therefore the space ${\cal X}_G(M)$ of generalized vector fields
together with the generalized commutator \{\;,\;\} form a Lie
algebra.

{\em S.C. would like to thank the organisers of the
DAE-BRNS High Energy Physics symposium, IIT KGP 2006 for the
invitation to present our work.}

\end{document}